# Achievable Rate Regions for the Dirty Multiple Access Channel with Partial Side Information at the Transmitters

Elham Bahmani and Ghosheh Abed Hodtani

Department of Electrical Engineering, Ferdowsi University of Mashhad, Mashhad, Iran

Bahmani.e@gmail.com, Ghodtani@gmail.com

*Abstract*— In this paper, we establish achievable rate regions for the multiple access channel (MAC) with side information partially known (estimated or sensed version) at the transmitters. Actually, we extend the lattice strategies used by Philosof-Zamir for the MAC with full side information at the transmitters to the partially known case. We show that the sensed or estimated side information reduces the rate regions, the same as that occurs for Costa Gaussian channel.

## I. INTRODUCTION

Nowadays channels with side information are widely studied from both information-theoretic and communications aspects.Side information (SI) can be available at the transmitter (SIT), and/or at the receiver (SIR). Encoding for a single-user with causal SIT was first studied by Shannon [1]. The capacity of a general discrete memoryless channel with non-causal SIT was characterized by Gel'fand and Pinsker in [2]. Costa [3] applied the formula obtained by Gel'fand and Pinsker to the special model of Gaussian channel with additive Gaussian interference, and showed that the channel capacity in the presence of interference known at the transmitter is the same as the case without interference. In Costa's dirty-paper channel (DPC), Gaussian random binning is able to eliminate the effect of interference which is known at the transmitter, and thus achieves capacity. Cover and Chiang [4] extended the above results and established a general capacity Theorem for the channel with two-sided state information. Gueguen-Sayrac [5] derived the capacity of the DPC with partial side information knowledge. The partial side information knowledge models the sensing process approximating the original information. It was shown that the capacity of the DPC with partial SI is reduced compared to the DPC with exact or complete SI.

In the multi-user setting, Das and Narayan [6] provided a multi-letter characterization of the capacity region of time-varying MACs with various degrees of SIT and SIR. In [7], a general framework for the capacity region of MACs with causal and non-causal SI was presented where focused on the MAC with independent side information at the two transmitters. Philosof-Zamir [8], [9], extended Jafar's work and provided achievable rate regions for the discrete memoryless MAC with correlated side information known non-causally at the encoders using a random binning technique. They also considered the Gaussian doubly dirty MAC in the high-SNR strong interference regime [8-13].The achievable rates using Costa's Gaussian binning vanish if both interference signals are strong. In contrast, it is shown that lattice-strategies (lattice pre-coding) can achieve positive rates, independent of the interference power. Furthermore, in some cases (which depend on the noise variance and power constraints) high dimensional lattice strategies are in fact optimal [10].

In this paper, we study the effect of partial SI knowledge in the Gaussian doubly dirty MAC considered by Philosof-Zamir [10]. We expect that achievable rate regions are reduced just the same as in Costa's DPC considered by in [5]. It is readily seen that our achievable rates include the achievable rates of the MAC with full side information as especial cases.

The rest of the paper is organized as follows. In Section II, we state some basic terminology for lattices. Section III includes related works. In Section IV, we state the system model and our results based on lattice strategies for doubly dirty MAC with partial SIT. In particular, we devote Section IV-A to the imbalanced doubly dirty MAC with partial SIT, and Section IV-B to the nearly balanced doubly dirty MAC with estimated SIT.

## II. LATTICES AND NESTED LATTICE CODES

We need some basic terminology for lattices before we can proceed and look at the modulo-lattice modulation. An $n$ dimensional lattice $\Lambda$ is defined by the generator matrix $G \in R^{n \times n}$. A point $l \in \mathbb{R}^n$ belongs to the lattice if and only if it can be written as $l = iG$, where $i \in \mathbb{Z}^n$ and $\mathbb{Z} = \{0, \pm 1, \pm 2, \dots\}$. The nearest neighbor quantizer of a lattice $\Lambda$ is defined by $Q_\Lambda(x) \triangleq \arg min_{l \in \Lambda} \|x - l\|$ where $\|.\|$ is the Euclidean norm. The modulo-lattice operation is defined by

$$x \, mod \, \Lambda = x - Q_\Lambda(x). \tag{1}$$

The modulo-code operation satisfies as follows

$$[\, x \, mod \, \Lambda + y\,] \, mod \, \Lambda = [\, x + y\,] \, mod \, \Lambda. \tag{2}$$

The fundamental Voronoi region of $\Lambda$ is the set of all points closer to the origin than to any other lattice point $\mathcal{V}(\Lambda) = \{x : Q(x) = 0\}$ with volume $V = \text{Vol}(\mathcal{V}(\Lambda))$. The second moment per dimension of a uniform distribution over $\mathcal{V}$ is

$$\sigma_\Lambda^2 = \frac{\frac{1}{n}\int_\mathcal{V} \|x\|^2 dx}{V} \, .$$

The normalized second moment is $G(\Lambda) = \frac{\sigma_\Lambda^2}{V^{2/n}}$. For large enough $n$, $\lim_n G(\Lambda) = \frac{1}{2\pi e}$, i.e., there exist good lattice quantizers and $\log(2\pi e G(\Lambda)) < \varepsilon$ for any $\varepsilon > 0, n \to \infty$. When used as a channel code over an unconstrained AWGN channel, [14], the decoding error probability is the probability that a white Gaussian noise vector $\mathbf{Z}$ exceeds the basic Voronoi cell $P_e = P_r(Z \notin \mathcal{V})$. For good AWGN channel coding, we have $P_e = P_r(Z \notin \mathcal{V}) < \varepsilon$ for any $\varepsilon > 0$. The Crypto lemma [15] which states that $(x + U) \mod \Lambda$ (where $U$ is uniformly distributed over $\mathcal{V}$) is an independent random variable uniformly distributed over $\mathcal{V}$.

The differential entropy of an $n$-dimensional random vector $\mathbf{D}$ which is distributed uniformly over the fundamental Voronoi cell, i.e., $\mathbf{D} \sim Unif(\mathcal{V})$ is given by [10]

$$h(\mathbf{D}) = \log_2(V) = \log_2\left(\frac{\sigma_\Lambda^2}{G(\Lambda)}\right)^{n/2}$$
$$= n/2 \log_2\left(\frac{\sigma_\Lambda^2}{G(\Lambda)}\right) = n/2 \log_2(2\pi e \sigma_\Lambda^2) \quad (3)$$

where the last (approximate) equality holds for lattices that are good for quantization. A comprehensive study of lattices and lattice quantization can be found in [16].

## III. RELATED WORKS

In this section, we briefly review the Gueguen-Sayrac and Philosof-Zamir works.

*Gueguen-Sayrac work:* Fig.1 shows the channel encoding with partial observation of the side information at the encoder. The partial observation is modeled as a compression channel generating some distortion on the side information. $\mathbf{X}$, $\mathbf{S}$, $\mathbf{Z}$ and $\mathbf{Y}$ are random variables which represent respectively, the source, side information, zero mean white Gaussian noise ($\mathbf{Z} \sim \mathcal{N}(0, N)$) and the output of the ergodic channel modeled by $\mathbf{Y} = \mathbf{X} + \mathbf{S} + \mathbf{Z}$. It is supposed that $\mathbf{X}$ and $\mathbf{S}$ are independent and average power limited variables $\mathbb{E}[\mathbf{X}^2] \leq P_X$ and $\mathbb{E}[\mathbf{S}^2] \leq P_s$.

$\tilde{\mathbf{S}}$ is the random variable representing the partial information obtained at the channel encoder which satisfies the following Euclidean distortion $\mathbb{E}\left[(\mathbf{S} - \tilde{\mathbf{S}})^2\right] = D$.

*Gueguen-Sayrac Theorem [5, Theorem 2.1]:* For a channel described as above, the channel capacity between the source and the output is expressed by $C = \frac{1}{2}\log\left(1 + \frac{P_X}{D + N}\right)$.

As we see the partial side information reduces the capacity.

*Philosof-Zamir work:* Fig.2 shows a general lattice-based transmission scheme as the Gaussian doubly dirty MAC. Encoder1 and encoder2 use the lattices $\Lambda_1 = k_1\Lambda$ and $\Lambda_2 = k_2\Lambda$ (where $k_1, k_2$ are real numbers), with second moments $\sigma_1^2 = P_1$ and $\sigma_2^2 = P_2$ and fundamental Voronoi regions $\mathcal{V}_1$ and $\mathcal{V}_2$, respectively. $\mathbf{V}_1 \in unif(\mathcal{V}_1)$, $\mathbf{V}_2 \in unif(\mathcal{V}_2)$ are independent and carry the information of user1 and user2. The encoders use independent (pseudorandom) dither signals $\mathbf{D}_1 \sim unif(\mathcal{V}_1)$ and $\mathbf{D}_2 \sim unif(\mathcal{V}_2)$ where $\mathbf{D}_1$ is known to encoder1 and to the decoder, and $\mathbf{D}_2$ is known to encoder2 and to the decoder. From the dithered quantization property, there is

$$\mathbf{X}_i \sim unif(\mathcal{V}_i) \text{ for any } \mathbf{V}_i = \mathbf{v}_i, i = 1,2 \quad (4)$$

where $\mathbf{X}_i$ independent of $\mathbf{V}_i$. $\mathbf{S}_i$, $i = 1,2$ is SI in the encoders. The transmitted signals by encoders are:

$$\mathbf{X}_1 = [\mathbf{V}_1 - \alpha_1 \mathbf{S}_1 + \mathbf{D}_1] mod \Lambda_1$$
$$\mathbf{X}_2 = [\mathbf{V}_2 - \alpha_2 \mathbf{S}_2 + \mathbf{D}_2] mod \Lambda_2 \quad (5)$$

where $\mathbf{X}_i$ independent of $\mathbf{V}_i$ and $\alpha_1, \alpha_2 \in [0,1]$. The input of decoder is $\mathbf{Y} = \mathbf{X}_1 + \mathbf{X}_2 + \mathbf{S}_1 + \mathbf{S}_2 + \mathbf{Z}$. The decoder uses a lattice $\Lambda_r = k_r \Lambda$, which is another scaled version of $\Lambda$, and reduces modulo-$\Lambda_r$ the term $\alpha_r \mathbf{Y} - \gamma \mathbf{D}_1 - \beta \mathbf{D}_2$, i.e.

$$\mathbf{Y}' = [\alpha_r \mathbf{Y} - \gamma \mathbf{D}_1 - \beta \mathbf{D}_2] mod \Lambda_r \quad (6)$$

The scalars $\alpha_1, \alpha_2, \alpha_r, k_1, k_2, k_r, \beta, \gamma$ and the basic lattice $\Lambda$ will be determined in each scenario. Robustness is the main advantage of the lattice-alignment transmission above [10]. They provide conditions under which lattice-strategies are optimal, imbalanced case $N \leq \sqrt{P_1 P_2} - \min(P_1, P_2)$ and nearly case $N \geq \sqrt{P_1 P_2} - \min(P_1, P_2)$.

*Philosof-Zamir Theorem [10, Theorem 2]:* The capacity region of the imbalanced doubly dirty MAC in the limit of strong interferences is given by the set of all rate pairs $(R_1, R_2)$ satisfying

$$R_1 + R_2 \leq \frac{1}{2} \log\left(1 + \frac{min(P_1, P_2)}{N}\right). \quad (7)$$

*Philosof-Zamir Theorem [10, Theorem 3]:* The inner bound of the capacity for the general nearly balanced case is given as follows

$$R_1 + R_2 \leq u.c.e\left\{\left[\frac{1}{2}\log\left(\frac{P_1 + P_2 + N}{2N + (\sqrt{P_1} - \sqrt{P_2})^2}\right)\right]^+\right\} \quad (8)$$

where the upper convex envelope is with respect to $P_1$ and $P_2$.

The inner bound for the exactly balanced case (nearly case with $P_1 = P_2$) is derived as follows

$$R_1 + R_2 \leq u.c.e\left\{\left[\frac{1}{2}\log\left(\frac{1}{2} + \frac{P}{N}\right)\right]^+\right\}. \quad (9)$$

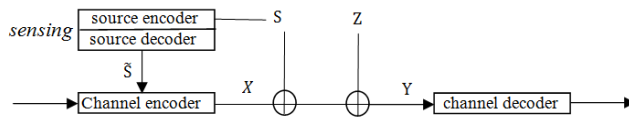

Fig.1 DPC with estimated SIT [5].

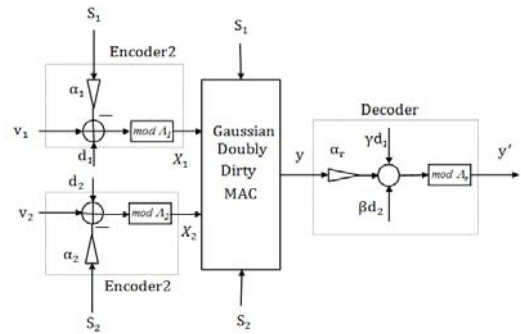

Fig.2 Gaussian Doubly Dirty MAC [10].

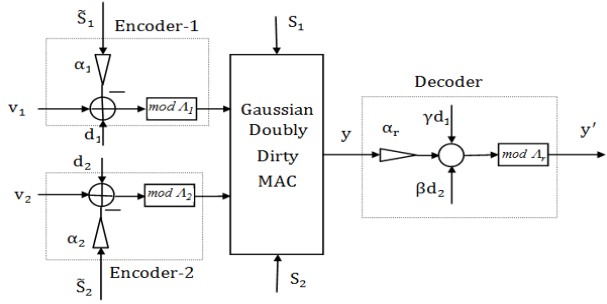

Fig.3 Gaussian Doubly Dirty MAC with partial SIT.

## IV. MAIN RESULTS

*System Model*

We consider the extended model of Gaussian doubly dirty MAC with partial side information in Fig.3. Most of the definitions and assumptions we make at this work are similar to [10], except that we have estimated SI ($\tilde{S}_i$) instead of exact SIT ($S_i$) at the transmitters which satisfies the following Euclidean distortion $\mathbb{E}\left[(S_i - \tilde{S}_i)^2\right] = E_i$, $i = 1,2$.

Therefore the transmitted signals by encoders are:

$X_1 = [V_1 - \alpha_1 \tilde{S}_1 + D_1] mod \Lambda_1$

$X_2 = [V_2 - \alpha_2 \tilde{S}_2 + D_2] mod \Lambda_2$ (10)

*Main Results:*

In the following Theorems, we provide conditions under which lattice-strategies are optimal (Like Philosof Zamir's work [10]). By considering two cases, imbalanced doubly dirty MAC with estimated (partial) SIT ($E_1 + E_2 + N \leq \sqrt{P_1 P_2} - \min(P_1, P_2)$) and nearly doubly dirty MAC with estimated (partial) SIT ($E_1 + E_2 + N \geq \sqrt{P_1 P_2} - \min(P_1, P_2)$), we give the achievable rate regions.

### A. Imbalanced Doubly Dirty MAC with partial SIT

We consider the case that $E_1 + E_2 + N \leq \sqrt{P_1 P_2} - \min(P_1, P_2)$ and named "imbalanced case".

**Theorem 1**: In imbalanced case for $P_1 \neq P_2$ the achievabile rate of the doubly dirty MAC with estimation of side information in transmitters is given by the set of all rate pairs $(R_1, R_2)$ satisfying

$R_1 + R_2 \leq \frac{1}{2} log \left(1 + \frac{\min(P_1, P_2)}{E_1 + E_2 + N}\right)$ (11)

**Corollary 1:** Theorem 1 is reduced to Theorem 2 [10] by $E_1 = E_2 = 0$.

**Proof of Theorem 1:** In this proof we consider four cases, the first case user1 is a helper for user2 where $P_1 \geq P_2 \left(\frac{P_2 + N + E_1 + E_2}{P_2}\right)^2$, the second case user1 is a helper for user2 where $P_2 \geq P_1 \left(\frac{P_1 + N + E_1 + E_2}{P_1}\right)^2$, the third case user2 is a helper for user1 where $P_1 \geq P_2 \left(\frac{P_2 + N + E_1 + E_2}{P_2}\right)^2$, the fourth case user2 is a helper for user1 where $P_2 \geq P_1 \left(\frac{P_1 + N + E_1 + E_2}{P_1}\right)^2$. Now, we show achievability for the first case i.e., for the point

$(R_1, R_2) \leq \left(0, \frac{1}{2} log \left(1 + \frac{\min(P_1, P_2)}{E_1 + E_2 + N}\right)\right).$ (12)

By applying the lattice transmission scheme and considering $\alpha_1 = \beta = k_1 = 1$, $k_2 = k_r = \gamma = \alpha_r = \alpha_2$, $V_1 = 0$ and $\sigma_1^2 = P_1$, $\sigma_2^2 = \alpha_2^2 P_1$, the encoder1 and encoder2 send $X_1$ and $X_2$, respectively, where they are generated as follows

$X_1 = [-\tilde{S}_1 + D_1] mod \Lambda_1$,

$X_2 = [V_2 - \alpha_2 \tilde{S}_2 + D_2] mod \Lambda_2$ (13)

The output of the decoder is $Y' = [\alpha_2(Y - D_1) - D_2] mod \Lambda_2$. Therefore we have

$Y' = [\alpha_2(X_1 + X_2 + S_1 + S_2 + Z - D_1) - D_2] mod \Lambda_2$ (14)
$= [\alpha_2 X_1 + X_2 - (1 - \alpha_2) X_2 + \alpha_2 S_1 + \alpha_2 S_2 + \alpha_2 Z - \alpha_2 D_1 - D_2] mod \Lambda_2$ (15)
$= [\alpha_2[-\tilde{S}_1 + D_1] mod \Lambda_1 + [V_2 - \alpha_2 \tilde{S}_2 + D_2] mod \Lambda_2 - (1 - \alpha_2) X_2 + \alpha_2 S_1 + \alpha_2 S_2 + \alpha_2 Z - \alpha_2 D_1 - D_2] mod \Lambda_2$ (16)
$= [V_2 - (1 - \alpha_2) X_2 + \alpha_2(S_1 - \tilde{S}_1) \alpha_2(S_2 - \tilde{S}_2) + \alpha_2 Z - \alpha_2 Q_{\Lambda_1}(-\tilde{S}_1 + D_1) +] mod \Lambda_2$ (17)
$= [V_2 - (1 - \alpha_2) X_2 + \alpha_2(S_1 - \tilde{S}_1) + \alpha_2(S_2 - \tilde{S}_2) + \alpha_2 Z] mod \Lambda_2$ (18)

where (16) follows from (13) and (17) follows from (1), (2). Since $\Lambda_1 = \Lambda$, $\Lambda_2 = \alpha_2 \Lambda$, and $\alpha_2 Q_{\Lambda_1}(-\tilde{S}_1 + D_1) \in \Lambda_2$, then this element disappears after the modulo-$\Lambda_2$ operation (in (18)). The rate achieved by user 2 is given by:

$R_2 = \frac{1}{n} I(V_2; Y') = \frac{1}{n} \{h(Y') - h(Y'|V_2)\}$ (19)
$= \frac{1}{n} \{h(Y') - h(-(1 - \alpha_2) X_2 + \alpha_2(S_1 - \tilde{S}_1) + \alpha_2(S_2 - \tilde{S}_1) + \alpha_2 Z)\}$ (20)
$\geq \frac{1}{2} log \left(\frac{P_2}{G(\Lambda_2)}\right) - \frac{1}{2} log \left(2\pi e \left((1 - \alpha_2)^2 P_2 + \alpha_2^2 (E_1 + E_2 + N)\right)\right)$ (21)
$= \frac{1}{2} log \left(\frac{P_2}{(1 - \alpha_2)^2 P_2 + \alpha_2^2 (E_1 + E_2 + N)}\right) - \frac{1}{2} log(2\pi e G(\Lambda_2))$ (22)

where (20) follows from this point that $V_2$ and $X_2$ are independent (according to (4)). Since $V_2$ is uniform over $\mathcal{V}_2$ then $Y'$ is uniform over $\mathcal{V}_2$. Gaussian distribution maximizes the entropy for fixed second moment and also modulo operation reduces the second moment, therefore we have inequality (21).

Using the optimal MMSE factor for user 2, we have

$\frac{\partial R_2}{\partial \alpha_2} = 0$, $\alpha_2^* = \frac{P_2}{P_2 + E_1 + E_2 + N}$. (23)

For $P_1 = P_2 \left(\frac{P_2 + N + E_1 + E_2}{P_2}\right)^2$ and for lattice that is good for quantization, the achievable rate is given by

$R_2 \leq \frac{1}{2} log \left(1 + \frac{P_2}{E_1 + E_2 + N}\right)$ (24)

Now we consider the second case (where $P_2 \geq P_1 \left(\frac{P_1 + N + E_1 + E_2}{P_1}\right)^2$) and show achievability for the point satisfying

$(R_1, R_2) = \left(0, \frac{1}{2} log \left(1 + \frac{P_1}{E_1 + E_2 + N}\right)\right)$ (25)

By applying the lattice transmission scheme and considering $\alpha_2 = \gamma = k_2 = 1$, $k_1 = k_r = \alpha_r = \alpha_1$, $\beta = 0$, $\mathbf{V}_1 = \mathbf{0}$, $\mathbf{D}_2 = \mathbf{0}$ and $\sigma_1^2 = \alpha_1^2 P_1$, $\sigma_2^2 = P_2$, the encoder1 and encoder2 send $\mathbf{X}_1$ and $\mathbf{X}_2$, respectively, where they are generated as follows

$$\mathbf{X}_1 = [-\alpha_1 \tilde{\mathbf{S}}_1 + \mathbf{D}_1] mod \Lambda_1 \ , \ \mathbf{X}_2 = [\mathbf{V}_2 - \tilde{\mathbf{S}}_2] mod \Lambda_2 \quad (26)$$

The receiver calculates $\mathbf{Y}' = [\alpha_1 \mathbf{Y} - \mathbf{D}_1] mod \Lambda_1$. By using (26), (1), (2) and this point that $\alpha_1 Q_{\Lambda_2}(\mathbf{V}_2 - \tilde{\mathbf{S}}_2) \in \Lambda_1$, the equivalent channel is given by

$$\mathbf{Y}' = [\alpha_1 \mathbf{V}_2 - (1 - \alpha_1)\mathbf{X}_1 + \alpha_1(\mathbf{S}_1 - \tilde{\mathbf{S}}_1) + \alpha_1(\mathbf{S}_2 - \tilde{\mathbf{S}}_2) + \alpha_1 \mathbf{Z}] mod \Lambda_1 \quad (27)$$

Since $\mathbf{V}_2$ and $\mathbf{X}_1$ are independent, the rate achieved by user 2 is given by

$$\mathbf{R}_2 = \tfrac{1}{n} I(\mathbf{V}_2; \mathbf{Y}') = \tfrac{1}{n}\{h(\mathbf{Y}') - h(\mathbf{Y}'|\mathbf{V}_2)\} \quad (28)$$
$$= \tfrac{1}{n}\{h(\mathbf{Y}') - h(-(1-\alpha_1)\mathbf{X}_1 + \alpha_1(\mathbf{S}_1 - \tilde{\mathbf{S}}_1) + \alpha_1(\mathbf{S}_2 - \tilde{\mathbf{S}}_2) + \alpha_1 \mathbf{Z})\}$$
$$\geq \tfrac{1}{2} log\left(\tfrac{P_1}{G(\Lambda_1)}\right) - \tfrac{1}{2} log\left(2\pi e\left((1-\alpha_1)^2 P_1 + \alpha_1^2(E_1 + E_2 + N)\right)\right)$$
$$= \tfrac{1}{2} log\left(\tfrac{P_1}{(1-\alpha_1)^2 P_1 + \alpha_1^2(E_1 + E_2 + N)}\right) - \tfrac{1}{2} log(2\pi e G(\Lambda_1)) \quad (29)$$

Since $\alpha_1 \mathbf{V}_2$ is uniform over $\mathcal{V}_1$ then $\mathbf{Y}'$ is also uniform over $\mathcal{V}_1$. Gaussian distribution maximizes the entropy for fixed second moment and also modulo operation reduces the second moment, therefore we have inequality (29). By using the optimal MMSE factor, we have

$$\alpha_1^* = \tfrac{P_1}{P_1 + E_1 + E_2 + N} \ . \quad (30)$$

For $P_2 = P_1 \left(\tfrac{P_1 + N + E_1 + E_2}{P_1}\right)^2$ and for lattice that is good for quantization, the rate is achieved as follows

$$R_2 \leq \tfrac{1}{2} log\left(1 + \tfrac{P_1}{E_1 + E_2 + N}\right). \quad (31)$$

Therefore, the achievable rate for the point $(0, R_2)$ satisfies

$$R_2 = \begin{cases} \tfrac{1}{2} log_2\left(1 + \tfrac{P_1}{N + E_1 + E_2}\right), & P_2 \geq P_1 \left(\tfrac{P_1 + N + E_1 + E_2}{P_1}\right)^2 \\ \tfrac{1}{2} log_2\left(1 + \tfrac{P_2}{N + E_1 + E_2}\right), & P_1 \geq P_2 \left(\tfrac{P_2 + N + E_1 + E_2}{P_2}\right)^2 \end{cases}. \quad (32)$$

By considering the third and fourth cases, also the point $(R_1, 0)$ is given by

$$R_1 = \begin{cases} \tfrac{1}{2} log_2\left(1 + \tfrac{P_1}{N + E_1 + E_2}\right), & P_2 \geq P_1 \left(\tfrac{P_1 + N + E_1 + E_2}{P_1}\right)^2 \\ \tfrac{1}{2} log_2\left(1 + \tfrac{P_2}{N + E_1 + E_2}\right), & P_1 \geq P_2 \left(\tfrac{P_2 + N + E_1 + E_2}{P_2}\right)^2 \end{cases}. \quad (33)$$

By using time sharing between (32) and (33) for $E_1 + E_2 + N \leq \sqrt{P_1 P_2} - min(P_1, P_2)$ and $P_1 \neq P_2$, any rate pair in the straight line $R_1 + R_2 \leq \tfrac{1}{2} log\left(1 + \tfrac{min(P_1, P_2)}{E_1 + E_2 + N}\right)$ is achievable and the Theorem follows.

In the above lattice-alignment scheme, the "strong user" (the user with higher power constraint) effectively uses $\alpha = 1$ (the scalar factor which multiplies the interference at the encoder).

*B. Nearly Balanced Doubly Dirty MAC with partial SIT*

We now derive an inner bound for the "nearly balanced" case, where $E_1 + E_2 + N \geq \sqrt{P_1 P_2} - min(P_1, P_2)$. For simplicity, we can divide nearly doubly dirty MACs with partial SIT to two cases, the symmetric ("exactly balanced") case, i.e., $P_1 = P_2 = P$ and the general "nearly balanced" case $P_1 \neq P_2$.

***Theorem 2:*** In exactly balanced case, the achievable rate of the doubly dirty MAC with estimated SIT is given by the set of all rate pairs $(R_1, R_2)$ satisfying

$$R_1 + R_2 \leq u.c.e\left\{\left[\tfrac{1}{2} log\left(\tfrac{1}{2} + \tfrac{P}{E_1 + E_2 + N}\right)\right]^+\right\} \quad (34)$$

***Corollary 2:*** Theorem 2 is reduced to (9) by $E_1 = E_2 = 0$.

***Proof of Theorem 2:*** Using the lattice-alignment transmission scheme, we consider the case that $k_1 = k_2 = k_r = \beta = \gamma = 1$ and $\alpha_1 = \alpha_2 = \alpha_r = \alpha$. The encoder1 and encoder2 send $\mathbf{X}_1$ and $\mathbf{X}_2$, respectively which are generated as follows

$$\mathbf{X}_1 = [\mathbf{V}_1 - \alpha \tilde{\mathbf{S}}_1 + \mathbf{D}_1] mod \Lambda, \ \mathbf{X}_2 = [\mathbf{V}_2 - \alpha \tilde{\mathbf{S}}_2 + \mathbf{D}_2] mod \Lambda \quad (35)$$

and the receiver calculates

$$\mathbf{Y}' = [\alpha \mathbf{Y} - \mathbf{D}_1 - \mathbf{D}_2] mod \Lambda \ . \quad (36)$$

By using (35) and (2), the equivalent channel is given by

$$\mathbf{Y}' = [\mathbf{V}_1 + \mathbf{V}_2 - (1 - \alpha)\mathbf{X}_1 - (1 - \alpha)\mathbf{X}_2 + \alpha(\mathbf{S}_1 - \tilde{\mathbf{S}}_1) + \alpha(\mathbf{S}_2 - \tilde{\mathbf{S}}_2) + \alpha \mathbf{Z}] mod \Lambda. \quad (37)$$

The sum of achievable rates is given by

$$R_1 + R_2 = \tfrac{1}{n} I(\mathbf{V}_1 \mathbf{V}_2; \mathbf{Y}') = \tfrac{1}{n}\{h(\mathbf{Y}') - h(\mathbf{Y}'|\mathbf{V}_1 \mathbf{V}_2)\} \quad (38)$$
$$= \tfrac{1}{n}\{h(\mathbf{Y}') - h([-(1-\alpha)\mathbf{X}_1 - (1-\alpha)\mathbf{X}_2 + \alpha(\mathbf{S}_1 - \tilde{\mathbf{S}}_1) + \alpha(\mathbf{S}_2 - \tilde{\mathbf{S}}_2) + \alpha \mathbf{Z}] mod \Lambda)\} \quad (39)$$
$$\geq \tfrac{1}{2} log\left(\tfrac{P}{G(\Lambda)}\right) - \tfrac{1}{2} log\left(2\pi e(2(1-\alpha)^2 P + \alpha^2(E_1 + E_2 + N))\right) \quad (40)$$
$$= \tfrac{1}{2} log\left(\tfrac{P}{2(1-\alpha)^2 P + \alpha^2(E_1 + E_2 + N)}\right) - \tfrac{1}{2} log(2\pi e G(\Lambda)). \quad (41)$$

By using the optimal MMSE factor, we achieve $\alpha^* = \tfrac{2P}{2P + E_1 + E_2 + N}$. Therefore, any rate pair satisfying (42) is achievable.

$$R_1 + R_2 \leq \left[\tfrac{1}{2} log\left(\tfrac{1}{2} + \tfrac{P}{E_1 + E_2 + N}\right)\right]^+. \quad (42)$$

Clearly, using a time sharing argument (34) can be achieved.

***Theorem 3:*** In the general "nearly balanced" case, where $E_1 + E_2 + N \geq \sqrt{P_1 P_2} - min(P_1, P_2)$. For general $P_1, P_2$, the achievable rate of the doubly dirty MAC with estimated SIT is given by the set of all rate pairs $(R_1, R_2)$ satisfying

$$R_1 + R_2 \leq u.c.e\left\{\left[\tfrac{1}{2} log\left(\tfrac{P_1 + P_2 + E_1 + E_2 + N}{2(E_1 + E_2 + N) + (\sqrt{P_1} - \sqrt{P_2})^2}\right)\right]^+\right\}. \quad (43)$$

***Corollary 3:*** Theorem 3 is reduced to (8) by $E_1 = E_2 = 0$.

***Proof of Theorem 3:*** For proof of this Theorem, we consider two cases $P_1 \leq P_2 \leq P_1 \left(\tfrac{P_1 + N + E_1 + E_2}{N + E_1 + E_2}\right)^2$ and $P_2 \leq P_1 \leq P_2 \left(\tfrac{P_2 + N + E_1 + E_2}{N + E_1 + E_2}\right)^2$. First, we consider the case of one and show achievability for the rate pair $(R_1, 0)$ where

$$R_1 = \frac{1}{2} \log \left( \frac{P_1+P_2+E_1+E_2+N}{2(E_1+E_2+N)+(\sqrt{P_1}-\sqrt{P_2})^2} \right). \quad (44)$$

By the use of the lattice transmission scheme, we consider $k_1 = k_r = \beta = \frac{\alpha_1}{\alpha_2}$, $k_2 = \gamma = 1$, $\alpha_1 = \alpha_r$, $\mathbf{V}_2 = \mathbf{0}$ and $\sigma_2^2 = P_2$, $\sigma_1^2 = \frac{\alpha_1^2}{\alpha_2^2} P_2$. The encoders send

$$\mathbf{X}_1 = [\mathbf{V}_1 - \alpha_1 \tilde{\mathbf{S}}_1 + \mathbf{D}_1] mod \Lambda_1, \quad \mathbf{X}_2 = [-\alpha_2 \tilde{\mathbf{S}}_2 + \mathbf{D}_2] mod \Lambda_2. \quad (45)$$

The receiver calculates $\mathbf{Y}' = [\alpha_1 \mathbf{Y} - \mathbf{D}_1 - \beta \mathbf{D}_2] mod \Lambda_1$. By using (45), (1), (2) and this point that $\frac{\alpha_1}{\alpha_2} Q_{\Lambda_2}(-\alpha_2 \tilde{\mathbf{S}}_2 + \mathbf{D}_2) \in \Lambda_1$, we have

$$\mathbf{Y}' = \left[ \mathbf{V}_1 - (1-\alpha_1)\mathbf{X}_1 - \frac{\alpha_1}{\alpha_2}(1-\alpha_2)\mathbf{X}_2 + \alpha_1 \mathbf{Z} + \alpha_1 (\mathbf{S}_1 - \tilde{\mathbf{S}}_1) + \alpha_1 (\mathbf{S}_2 - \tilde{\mathbf{S}}_2) \right] mod \Lambda_1 \quad (46)$$

Since $\mathbf{X}_1$ is independent of $\mathbf{V}_1$ and $\mathbf{X}_2$ is independent of $\mathbf{V}_1$ and $\mathbf{X}_1$, then we can obtain $R_1$ as

$$R_1 = \frac{1}{n} I(\mathbf{V}_1; \mathbf{Y}') = \frac{1}{n} \{ h(\mathbf{Y}') - h(\mathbf{Y}'|\mathbf{V}_1) \}$$
$$= \frac{1}{n} \left\{ h(\mathbf{Y}') - h \left( \left[ -(1-\alpha_1)\mathbf{X}_1 - \frac{\alpha_1}{\alpha_2}(1-\alpha_2)\mathbf{X}_2 + \alpha_1 \mathbf{Z} + \alpha_1(\mathbf{S}_1 - \tilde{\mathbf{S}}_1) + \alpha_1(\mathbf{S}_2 - \tilde{\mathbf{S}}_2) \right] mod \Lambda_1 \right) \right\} \quad (47)$$

$$\geq \frac{1}{2}\log\left(\frac{P_1}{G(\Lambda_1)}\right) - \frac{1}{2}\log\left(2\pi e\left((1-\alpha_1)^2 P_1 + \left(\frac{\alpha_1}{\alpha_2}\right)^2 (1-\alpha_2)^2 P_2 + \alpha_1^2 (E_1 + E_2 + N)\right)\right) \quad (48)$$

By considering $\frac{\alpha_1}{\alpha_2} = \sqrt{\frac{P_1}{P_2}}$ and $G(\Lambda_1) \to \frac{1}{2\pi e}$ as $n \to \infty$ (good lattice), and by using the optimal MMSE factor, we have $\alpha_1^* = \frac{\sqrt{P_1}(\sqrt{P_1}+\sqrt{P_2})}{P_1+P_2+(E_1+E_2+N)}$. Therefore the rate region is achieved.

Now the second case is considered i.e, $P_2 \leq P_1 \leq P_2 \left( \frac{P_2+N+E_1+E_2}{N+E_1+E_2} \right)^2$. Again we show achievability for the rate pair $(R_1, 0)$. By the use of the lattice transmission scheme, we consider $k_2 = k_r = \gamma = \frac{\alpha_2}{\alpha_1}$, $k_1 = \beta = 1$, $\alpha_2 = \alpha_r$, $V_2 = \mathbf{0}$ and $\sigma_2^2 = P_2$, $\sigma_1^2 = \frac{\alpha_1^2}{\alpha_2^2} P_2$. The encoder1 and encoder2 send $\mathbf{X}_1$ and $\mathbf{X}_2$, respectively which are generated as follows

$$\mathbf{X}_1 = [\mathbf{V}_1 - \alpha_1 \tilde{\mathbf{S}}_1 + \mathbf{D}_1] mod \Lambda_1, \quad \mathbf{X}_2 = [-\alpha_2 \tilde{\mathbf{S}}_2 + \mathbf{D}_2] mod \Lambda_2 \quad (49)$$

and the receiver calculates $\mathbf{Y}' = [\alpha_2 \mathbf{Y} - \gamma \mathbf{D}_1 - \mathbf{D}_2] mod \Lambda_2$. Similar to first case, we achieve

$$R_1 = \frac{1}{2} \log \left( \frac{P_1+P_2+E_1+E_2+N}{2(E_1+E_2+N)+(\sqrt{P_1}-\sqrt{P_2})^2} \right). \quad (50)$$

Therefore, the achievable rate of the point $(R_1, 0)$ for $E_1 + E_2 + N \geq \sqrt{P_1 P_2} - min(P_1, P_2)$ is given by

$$(R_1, 0) = \left( \left[ \frac{1}{2} \log \left( \frac{P_1+P_2+E_1+E_2+N}{2(E_1+E_2+N)+(\sqrt{P_1}-\sqrt{P_2})^2} \right) \right]^+, 0 \right). \quad (51)$$

Due to the symmetry, it can be shown that the achievable rate of the point $(0, R_2)$ for $E_1 + E_2 + N \geq \sqrt{P_1 P_2} - min(P_1, P_2)$ is given by

$$(0, R_2) = \left( 0, \left[ \frac{1}{2} \log \left( \frac{P_1+P_2+E_1+E_2+N}{2(E_1+E_2+N)+(\sqrt{P_1}-\sqrt{P_2})^2} \right) \right]^+ \right). \quad (52)$$

By using a time sharing between the achievable rate pairs in (51) and (52), the proof of Theorem 3 is completed.

We can see that the result of Theorem 2 is achieved by considering $P_1 = P_2 = P$.